# Simulations of Pedestrian Impact Collisions with Virtual CRASH 3 and Comparisons with IPTM Staged Tests

Tony Becker, ACTAR, Mike Reade, Bob Scurlock, Ph.D., ACTAR

**Introduction**

In this article, we present results from a series of Virtual CRASH-based pedestrian impact simulations. We compare the results of these Virtual CRASH pedestrian impact simulations to data from pedestrian impact collisions staged at the Institute of Police Technology and Management.

**Staged Pedestrian Impact Experiments**

Each year the Institute of Police Technology and Management (IPTM) stages a series of pedestrian impact experiments as a part of its Pedestrian and Bicycle Crash Investigation courses. These experiments offer a unique opportunity for participants to gain hands-on experience setting up controlled experiments, gathering data and evidence, and performing full analyses of the collision events using standard reconstruction approaches in the accident reconstruction community. These analyses can then be compared to direct measurements obtained during the experiments.

**The Virtual CRASH Simulator**

Virtual CRASH 3 is a general-purpose fully three-dimensional accident reconstruction software package developed by Virtual CRASH, s.r.o., a company based out of Slovakia. Virtual CRASH is a simulation package that uses rigid-body dynamics to simulate collisions between vehicle objects within its environment; Virtual CRASH simulates multibody collisions in a manner similar to packages such as MADYMO [1] or Articulated Total Body [2]. The impact dynamics are also determined by pre-impact geometry and specification of the coefficients-of-friction and -restitution, as well as the inertial properties of the objects, which are all specified in the user interface. Virtual CRASH offers a unique, fast, and visually appealing way to simulate pedestrian impacts.

**Sanity Checks**

*Simulation of Dissipative Forces*

To better understand how Virtual CRASH performs compared to our expectations from classical physics, we conducted a series of ground slide and projectile motion experiments for a cylindrical "puck" and a dummy model. The puck was given the same mass as the default pedestrian model.

First, we evaluated the distances required for the puck and dummy systems to slide to a stop via frictional forces as they traveled along level flat terrain. In our treatment below, we mathematically model the puck and dummy systems as point-like particles. From the work-energy theorem [3], we expect a change in kinetic energy to be associated with dissipative frictional forces. That is:

$$W = \int_i^f \bar{F} \cdot d\bar{r} = \Delta KE$$
$$= \frac{1}{2}mv_f^2 - \frac{1}{2}mv_i^2 \qquad (1)$$

where the mass is displaced from point *i* to point *f* along some path. In the one-dimensional constant frictional force approximation, we have:

$$\bar{F} = -\mu m g \hat{x}$$

and

$$d\bar{r} = d\bar{x}$$

where $\hat{x}$ points along the direction of displacement. Assuming the mass comes to rest at point *f*, we can write an expression relating the total sliding distance to the pre-slide velocity at point *i*, $v_i$. This is given by:

$$-\mu m g \Delta x = -\frac{1}{2}mv_i^2$$

or more familiarly:

$$\Delta x = \frac{v_i^2}{2g\mu} \qquad (2)$$

where $\Delta x = x_f - x_i$. We conducted a series of experiments in Virtual CRASH, testing the distance required to stop a mass as a function of pre-slide speed. The puck and dummy objects started at ground height and were given an initial horizontal velocity. Rearranging (2), one expects the simulation to yield results consistent with the relation:

$$v_i^2 = \mu \cdot (2g\Delta x) \qquad (3)$$

Figure 1 illustrates the square of the initial velocity, $v_i^2$, plotted as a function of the quantity $(2g\Delta x)$. In the simulations, the puck was given an initial velocity between 10 to 50 mph. The coefficient-of-friction was varied from 0.1 to 1.0 and the total sliding distance was noted. As indicated by equation (3), the slope of a first-order polynomial fit to each set of points should yield the coefficient-of-friction used in the corresponding simulations. Excellent agreement is observed between the analytic solution and the simulation as indicated by the slopes of the fits, which serve as estimates of μ.

Figure 2 illustrates the corresponding results for the same experiments, where the Virtual CRASH default dummy was used rather than the puck. Again we see excellent agreement between the analytic solution and simulation, with only some slight deviation at high friction values. This is due to the dummy's body rotating toward the end of its motion at high speeds.

Figure 3 illustrates the sliding distance and speed versus time behavior of the puck and dummy, where each is given an initial speed of 30 mph, and a ground contact coefficient-of-friction of 0.6 is used in the simulations.[1] We see very good agreement in the behaviors of the puck and dummy. Indeed, focusing on the velocity versus time behavior of the two systems in Figure 4, first-order polynomial fits are performed to both sets of data. With a simulated ground contact coefficient-of-friction value of 0.6, the puck's average deceleration rate is within 0.3% of the expected value, and the dummy's average deceleration rate is within 0.7% of the expected value.

*Airborne Trajectory Simulation*

Virtual CRASH is a fully three-dimensional simulation environment; therefore, collision forces can project objects vertically, causing them to go airborne for some time. Indeed, the user can run simulations specifying an arbitrary initial velocity vector orientation at the start of simulation, allowing objects to go airborne independent of collision events.

For clarity, let the *x*-direction point along the forward direction. Let the *z*-direction point upward vertically, antiparallel to the gravitation acceleration direction. In the following, we will neglect restitution effects, and so assume that the projectiles either come to rest upon landing or slide to rest at some time after initial ground contact. From basic kinematics, for a point-mass object with initial height at ground level, launched at a velocity of magnitude, $|\bar{v}_i|$, at an angle $\theta$ with respect to the *x*-axis, we expect the *z*-position as a function of time to be given by:

$$z(t) = v_{z,i}t - \frac{1}{2}gt^2$$
$$= |\bar{v}_i|\sin(\theta)t - \frac{1}{2}gt^2 \qquad (4)$$

The total time, $t_a$, in which the object is airborne can be solved for by obtaining the solution to the quadratic:

$$z(t_a) = |\bar{v}_i|\sin(\theta)t_a - \frac{1}{2}gt_a^2 = 0 \qquad (5)$$

which yields:

---

[1] See video online at: https://youtu.be/htIwFYLG_W8

$$t_a = \frac{2|\bar{v}_i|\sin(\theta)}{g} \quad (6)$$

Figure 5 shows the total airborne time for simulated 30 mph launches of the puck and dummy systems as a function of $\sin(\theta)$.[2] First-order polynomial fits to these data points yield slopes that estimate the quantity $2|\bar{v}_i|/g$. Excellent agreement is observed between our analytic solution and simulation to the 1/100% level for both systems.

Since the object undergoes uniform motion along the horizontal direction (neglecting air resistance), we can solve for the total horizontal airborne distance by:

$$D_a = v_{x,i} t_a = |\bar{v}_i|\cos(\theta) t_a$$
$$= \frac{2|\bar{v}_i|^2 \sin(\theta) \cdot \cos(\theta)}{g} \quad (7)$$

Equation (7), of course, is the famous range equation from classical physics.

Figure 6 illustrates the total airborne distance as a function of the quantity $(\sin(\theta) \cdot \cos(\theta))$ for 30 mph launches of the puck and dummy systems. First-order polynomial fits to the data serve to estimate the quantity $(2|\bar{v}_i|^2/g)$. Again, excellent agreement between the simulations and the analytic solution is observed to better than a fraction of a percent.

*Ground Contact*

During the landing phase of projectile motion, the test mass interacts with the ground such that a contact-force is imparted to the mass, $\bar{F}(t)$, over a time $\Delta t$, which arrests the vertical motion of the mass and simultaneously retards its horizontal velocity. The impulse imparted to the test mass is given by [4]:

$$\bar{J} = \int_0^{\Delta t} dt \cdot \bar{F} = m \Delta \bar{v}$$
$$= \int_0^{\Delta t} dt \cdot F_n \hat{n} + \int_0^{\Delta t} dt \cdot F_t \hat{t} \quad (8)$$

where $\hat{n}$ is the unit vector pointing along the direction normal to the surface of contact (ground), and $\hat{t}$ points in the direction orthogonal to $\hat{n}$. Along the normal-direction we have:

$$J_n = \int_0^{\Delta t} dt \cdot F_n = m \Delta v_n \quad (9)$$

and along the tangent direction, we have:

$$J_t = \int_0^{\Delta t} dt \cdot F_t = m \Delta v_t \quad (10)$$

where $\Delta v_n$ and $\Delta v_t$ are the normal and tangent axis projections of change-in-velocity vector $\Delta \bar{v}$ respectively. The "impulse ratio" is given by:

$$\mu = \frac{J_t}{J_n} \quad (11)$$

---

[2] A video of the 30 mph 40 degree launch is online at: https://youtu.be/3aSFtCC6y4c

---

Therefore, from (9), (10), and (11) we have:

$$\Delta v_t = \mu \cdot \Delta v_n \quad (12)$$

The sign of the impulse ratio is determined by the relative velocity vector component tangent to the contact surface at the moment of contact. The impulse ratio is typically associated with the inter-object surface contact coefficient-of-friction, whose average behavior is often referred to as the *drag factor*. In this paper, we use the terms interchangeably since we neglect any time-dependent behavior of $\mu$.

The coefficient-of-restitution is given by the ratio of normal components of the final to initial relative velocities at the point-of-contact.

$$\varepsilon = -\frac{v_{Reln,f}}{v_{Reln,i}} \quad (13)$$

where the relative velocity is defined as the difference between the velocity vectors of the two interacting objects at the point-of-contact:

$$\bar{v}_{Rel} = \bar{v}_1 - \bar{v}_2 \quad (14)$$

The normal and tangent projections of this are given by:

$$v_{Reln} = \bar{v}_{Rel} \cdot \hat{n} \quad (15)$$

and

$$v_{Relt} = \bar{v}_{Rel} \cdot \hat{t} \quad (16)$$

Finally, we can rewrite the normal change-in-velocity as:

$$\Delta v_{Reln} = v_{Reln,f} - v_{Reln,i}$$
$$= -\varepsilon \cdot v_{Reln,i} - v_{Reln,i}$$

or

$$\Delta v_{Reln} = -(1+\varepsilon) \cdot v_{Reln,i} \quad (17)$$

and

$$\Delta v_{Relt} = -\mu(1+\varepsilon) \cdot v_{Reln,i} \quad (18)$$

Again, we treat the objects as point-like masses for our simplified mathematical model. For two objects undergoing a collision, the change-in-relative-velocity is related to the change-in-velocity at the center-of-gravity of object 1 is given by the relation [4]:

$$\Delta \bar{v}_1 = \left(\frac{\bar{m}}{m_1}\right) \Delta \bar{v}_{Rel} \quad (19)$$

Here, the system's reduced mass is given by:

$$\bar{m} = \frac{m_1 \cdot m_2}{m_1 + m_2} \quad (20)$$

In the limit where the mass of object 2 becomes infinite (such in a ground impact), we have the following:

$$\Delta \bar{v}_1 = \Delta \bar{v}_{Rel} \quad (21)$$

---

$$\Delta v_n = -(1+\varepsilon) \cdot v_{Reln,i} \quad (22)$$

$$\Delta v_t = -\mu(1+\varepsilon) \cdot v_{Reln,i} \quad (23)$$

Let us assume the ground is a flat level surface such $\hat{n} = \hat{z}$ and $\hat{t} = \hat{x}$. Let object 1 be an object undergoing projectile motion, and let object 2 be the infinitely massive ground plane. In the no-restitution limit, we have at time $t_a + \Delta t$:

$$\Delta v_z = |v_z(t_a)| \quad (24)$$

and

$$\Delta v_x = -\mu \cdot |v_z(t_a)| \quad (25)$$

where the collision pulse width, $\Delta t$, can be taken as vanishingly small. Figure 7 illustrates the relation between the simulated horizontal changes-in-velocity from and vertical changes-in-velocity for the puck and dummy systems at the moment just after ground impact, after the downward vertical velocity has been arrested. The plots in this figure were created using 30 mph launch speeds at increasing launch angles between 5 degrees (lowest $\Delta v_z$) to 85 degrees (largest $\Delta v_z$) from horizontal. There are a few interesting features of note in this figure. First, from equation (25), we expect a first-order polynomial fit to yield a slope equal to the ground contact coefficient-of-friction used in the simulations. This is indeed the case to within 0.4% for the puck system and 6% for the dummy model. We also note in the dummy model simulations, when $\Delta v_z$ exceeds 15 mph (30 degrees), $\Delta v_x$ deviates from its initial linear behavior. For $\Delta v_z > 25$ mph we see a dramatic drop in $\Delta v_x$ for the puck system.

For the dummy system, there are two effects that explain the non-linear behavior. First, as the launch angle increases, thereby increasing $\Delta v_z$, the torque imparted to the dummy's body increases, causing an increase in rotational kinetic energy rather than a decrease in linear kinetic energy. Therefore, we see a slight reduction in the expected horizontal change-in-velocity. In the second region, $\Delta v_z > 25$ mph, we see the same sharp reduction in $\Delta v_x$ as in the puck system. This is related to the reduction in the maximum possible horizontal impulse that can be delivered to the objects, which is naturally bounded such that the maximum horizontal change-in-velocity cannot exceed the initial horizontal launch speed. This is further discussed below.

*Post-Ground Contact*

The vertical velocity component at first ground contact is given by:

$$v_{z,f} = v_{z,i} - gt_a$$
$$= |\bar{v}_i|\sin(\theta) - g \cdot \frac{2|\bar{v}_i|\sin(\theta)}{g} \quad (26)$$

which simplifies to:

$$v_{z,f} = -|\bar{v}_i|\sin(\theta) \quad (27)$$

With this and equation (25), we can now solve for the final ground speed after landing. This is given by:

$$v_{x,f} = v_{x,i} + \Delta v_x$$
$$= |\bar{v}_i|\cos(\theta) - \mu|\bar{v}_i|\sin(\theta)$$
$$= |\bar{v}_i| \cdot (\cos(\theta) - \mu \cdot \sin(\theta)) \quad (28)$$

Assuming kinetic energy is dissipated through ground-contact frictional forces, we can use equation (2) to solve for the total slide distance.

$$D_{Slide} = \frac{v_{x,f}^2}{2g\mu}$$
$$= \frac{|\bar{v}_i|^2}{2g\mu} \cdot (\cos(\theta) - \mu \cdot \sin(\theta))^2 \quad (29)$$

The total projectile travel distance is given by the sum of the sliding distance and the airborne travel distance; that is:

$$D_{Total} = D_a + D_{Slide} \quad (30)$$

or

$$D_{Total} = \frac{2|\bar{v}_i|^2 \sin(\theta) \cdot \cos(\theta)}{g} + \frac{|\bar{v}_i|^2 \cdot (\cos(\theta) - \mu \cdot \sin(\theta))^2}{2g\mu} \quad (31)$$

which simplifies to:

$$D_{Total} = \frac{|\bar{v}_i|^2}{2g\mu} \cdot (\cos(\theta) + \mu \cdot \sin(\theta))^2 \quad (32)$$

Figure 8 illustrates the total throw distance as a function of the quantity $(\cos(\theta) + \mu \cdot \sin(\theta))^2$ for the puck and dummy systems with 30 mph launch speeds and $\mu = 0.6$. The slope of a first-order polynomial fit yields estimates of the quantity $(|\bar{v}_i|^2/2g\mu)$. This estimate is within 0.05% of the expected value for the puck and within 0.6% for the dummy. Again, we see the deviation from the linear behavior for higher angles that is associated with the expected reduction of horizontal impulse at large launch angles. This is explored below.

Here we note that solving the above expression for $|\bar{v}_i|$ yields the familiar *Searle Equation*:

$$|\bar{v}_i| = \frac{\sqrt{2g\mu D_{Total}}}{\cos(\theta) + \mu \cdot \sin(\theta)} \quad (33)$$

Returning to equation (32) above, taking the first derivative gives:

$$\frac{\partial D_{Total}}{\partial \theta} = \frac{|\bar{v}_i|^2}{g\mu} \cdot (\cos(\theta) + \mu \cdot \sin(\theta))$$
$$\times (-\sin(\theta) + \mu \cdot \cos(\theta)) \quad (34)$$

Solving for the extremum values requires the following relation to hold:

$$-\sin(\theta') + \mu \cdot \cos(\theta') = 0 \quad (35)$$

whose solution is given by:

$$\tan(\theta') = \mu \quad (36)$$

To simplify the above expression, let:

$$\cos(\theta) = \frac{\tilde{x}}{\sqrt{\tilde{x}^2 + \tilde{y}^2}} \quad (37)$$

and

$$\sin(\theta) = \frac{\tilde{y}}{\sqrt{\tilde{x}^2 + \tilde{y}^2}} \quad (38)$$

Therefore, using (37) and (38), we have:

$$\cos(\theta) + \mu \cdot \sin(\theta) = \frac{\tilde{x} + \mu\tilde{y}}{\sqrt{\tilde{x}^2 + \tilde{y}^2}} \quad (39)$$

The extremum condition above is now given by:

$$-\tilde{x}' + \mu \cdot \tilde{y}' = 0 \quad (40)$$

or,

$$\cos(\theta') + \mu \cdot \sin(\theta') =$$
$$\frac{\tilde{x}' \cdot (1 + \mu^2)}{\tilde{x}'\sqrt{1 + \mu^2}} = \sqrt{1 + \mu^2} \quad (41)$$

Therefore, using (41), the total throw distance at angle $\theta'$, is given by:

$$D_{Total}(\theta') = \frac{|\bar{v}_i|^2}{2g\mu} \cdot \left(\sqrt{1 + \mu^2}\right)^2$$
$$= \frac{|\bar{v}_i|^2}{2g\mu} \cdot (1 + \mu^2) \quad (42)$$

Checking the concavity at $\theta'$, we apply the second derivative:

$$\frac{\partial^2 D_{Total}}{\partial^2 \theta} = \frac{v_i^2}{g\mu} \times \{(-\sin(\theta') + \mu \cdot \cos(\theta'))$$
$$\times (-\sin(\theta') + \mu \cdot \cos(\theta'))$$
$$+ (\cos(\theta') + \mu \cdot \sin(\theta'))$$
$$+ (\cos(\theta') + \mu \cdot \sin(\theta'))$$
$$\times (-\cos(\theta') - \mu \cdot \sin(\theta'))\} \quad (43)$$

which simplifies to:

$$\frac{\partial^2 D_{Total}}{\partial^2 \theta}(\theta') = -\frac{|\bar{v}_i|^2}{g\mu\sqrt{1 + \mu^2}} < 0 \quad (44)$$

Thus, the second derivative is negative definite, implying that our function $D_{Total}(\theta)$ at $\theta'$ is indeed a maximum value.

*Limit on Horizontal Impulse*

Let us now focus on the particular behavior of our test mass just after ground impact. We know a retarding impulse is imparted to our mass upon ground impact; this is given by equation (10). This tangent impulse will only be applied so long as there is relative motion along the tangent axis direction between the interacting objects at the point of contact. Once the relative motion vanishes, this impulse component no longer acts on the mass. Therefore, the following relation must hold true for our expression for $D_{Total}(\theta)$ given by equation (32) to remain valid:

$$v_{x,f} = |\bar{v}_i| \cdot (\cos(\theta) - \mu \cdot \sin(\theta)) \geq 0 \quad (45)$$

This implies the condition:

$$\cos(\theta) - \mu \cdot \sin(\theta) \geq 0 \quad (46)$$

There are two equivalent ways to interpret this condition. First, we can solve for the angle, which gives the solution to $\cos(\theta) - \mu \cdot \sin(\theta) = 0$. This gives us a *boundary angle*:

$$\tilde{\theta} = \tan^{-1}\left(\frac{1}{\mu}\right) \quad (47)$$

Therefore, when the launch angle satisfies the condition $\theta < \tilde{\theta}$, equation (32) holds. Otherwise, if the condition is violated, the total distance is simply given by the range equation (7), where no sliding is expected. Thus, we have:

$$D_{Total} = \begin{cases} \frac{|\bar{v}_i|^2 \cdot (\cos(\theta) + \mu \cdot \sin(\theta))^2}{2g\mu}, & \theta < \tilde{\theta} \\ \frac{2|\bar{v}_i|^2 \sin(\theta) \cdot \cos(\theta)}{g}, & \theta \geq \tilde{\theta} \end{cases} \quad (48)$$

Figure 9 shows the total throw distance as function of launch angle for 30 mph launches with $\mu = 0.6$ for both puck and dummy systems. We see the simulated results track very closely to our analytic solutions. We also see beyond the calculated boundary angle at 59 degrees, the simulated behavior switches from following the Searle equation to the Range equation. Table 1 shows the difference between the analytic solution and simulated results. The simulated puck system shows better than 0.4% agreement and the dummy system shows better than 4% agreement with the analytic solution given by equation (48), each with much lower averages. This is shown in Table 1.

We note here that the equation for $\tilde{\theta}$ has a dependence on restitution that has been neglected in this treatment and therefore will have different behavior when generalized to account for this effect. This will be the topic of a future article.

The second way to interpret equation (46), is as an upper limit on the impulse ratio.

$$\mu(\theta) \leq 1/\tan(\theta) \quad (49)$$

That is, when using the Searle equation, one could use the following form for the coefficient-of-friction:

$$\mu = \min\left(f_d, \frac{1}{\tan(\theta)}\right) \quad (50)$$

where $f_d$ is the measured or typical ground contact drag factor for the subject case. Let us suppose we know the launch angle is sufficiently large such that we must use

$$\mu(\theta) = 1/\tan(\theta) \quad (51)$$

Substituting (51) into equation (32) gives:

$$D_{Total} = \frac{|\bar{v}_i|^2}{2g(1/\tan(\theta))}$$
$$\times (\cos(\theta) + (1/\tan(\theta)) \cdot \sin(\theta))^2$$
$$= \frac{2|\bar{v}_i|^2 \sin(\theta) \cdot \cos(\theta)}{g} \quad (52)$$

which is simply equation (7), thereby indicating that when the condition of no post-impact ground speed occurs, the total throw distance is simply given by the classical range equation, as expected.

Figure 10 shows the launch speed as a function of total throw distance for simulated launches between 10 and 50 mph, held at 20 degree launch angles. The coefficient-of-friction is set to 0.6. The corresponding Searle curve from equation (33) is shown as well. The Searle minimum and maximum curves are also drawn. Good agreement is evident between the Searle equation and the simulated behavior as shown in Table 2. Note there is no deviation from Searle behavior, as the 20 degree launch angle is well below the 59 degree boundary angle given by equation (47).

**Staged Collisions at IPTM**

On August 12, 2015, a series of staged impacts were conducted at the Institute of Police Technology and Management at the University of North Florida. These impacts were conducted as a part of the IPTM course on Pedestrian and Bicycle Crash Investigation. An aerial view of the test site can be seen in Figure 11. During these experiments, a plastic anthropomorphic dummy was impacted by a 2005 Ford Crown Victoria. The dummy was 49 lbs in weight and measured 5.83 feet in height. Prior to impact, the dummy was suspended in an upright position, using high-strength fishing line attached to a boom (Figure 12). The boom was mounted to the rear of a truck offset from the collision path. Upon impact, the fishing line broke free of the boom, allowing the dummy to effectively interact with the vehicle structure, unimpeded. High-speed video footage was captured of each collision event. Measurements of the post-impact travel distance were recorded after each impact, as well as the average deceleration rate of the 2005 Ford Crown Victoria, which was made to hard brake upon impact. The point of the dummy's first ground contact was also carefully recorded (Figure 13). The impact speed was also recorded for each test. A summary of the test data is given in Table 3 and Figure 14. Participants of the 40 hour program conducted an accident reconstruction analysis of each impact, based on the total throw distance of the dummy. Reconstructed pre-impact vehicle speed estimates were compared to known values recorded during each collision.

*Comparison of Staged Collision Data with Virtual CRASH Dummy Behavior*

A series of simulations were run in the Virtual CRASH 3 software environment. The simulated dummy's height and weight were set to match that of the IPTM test dummy described above. The simulated Crown Victoria passenger vehicle weight was set equal to that of the IPTM test vehicle weight plus driver weight. The simulated ground-contact coefficient-of-friction value was set to the average measured at the scene (0.511), as was the vehicle-contact coefficient-of-friction (0.5). A coefficient-of-restitution of 0 was used for both ground and vehicle contact. Table 4 gives a summary of parameters used for the Virtual CRASH simulation runs.

The impact location along the Ford's front bumper was set to either 0.5 ft or 1.5 ft away from the center line, as this was not a well-controlled parameter during the staged IPTM tests. The simulated dummy's pre-impact orientation was set to either 0 degrees (impact to rear of dummy) or 90 degrees (impact to left side of dummy). Figure 15 depicts the moment-of-impact for two of the simulated scenarios. It was found that there were no significant differences in the throw distances for impacts to the simulated dummy's left side compared to right side.

From (48), we expect that the simulation should yield the following relation:

$$|\bar{v}_i| \propto \sqrt{D_{Total}} \qquad (53)$$

Figure 16 illustrates the dependence of impact speed as a function of $\sqrt{D_{Total}}$. First-order polynomial fits are shown as well. Indeed, we see linear behavior over the ensemble of the test runs for each given test scenario. Detailed summaries of the results are shown in Table 5 and Table 6. To quantify how well Virtual CRASH simulates the dummy behavior observed in the IPTM staged tests, we calculate the differences (residuals) between first-order polynomial fits to the Virtual CRASH results and the test data. A plot of the residuals is shown in Figure 17. Here we see that the best match to set of IPTM data is given by Scenario 3, where the maximum deviation of the predicted impact speed based on total throw distance using Virtual CRASH simulations is less than 3 mph when compared to the IPTM dataset.

Figure 18 depicts data from Scenarios 1 and 3, as well as the Searle Minimum curve. The IPTM dataset is shown. We also show the best fit to data for wrap trajectory impacts aggregated in the meta-analysis presented by Happer et al., along with the corresponding 85% prediction interval [5]. Here we see excellent agreement between Virtual CRASH simulated impacts, IPTM data, and expectations from prior studies.

*Simulating Gross Behavior of Test Dummy*

In addition to simulating the total throw versus impact speed behavior of the test data, we wanted to see if we could simulate the overall behavior of the test dummy's motion using Virtual CRASH. The height and weight of the crash test dummy were both input into Virtual CRASH. The joint stiffness properties of the Virtual CRASH simulated dummy can be adjusted such that the user can tune the dummy's overall rigidity. This can be done separately for each joint if needed. We chose to focus on adjusting three of the dummy parameters to get reasonable agreement between behavior observed in the staged collision 4 video and the Virtual CRASH output: these parameters were the overall coefficient-of-restitution, coefficient-of-friction, and joint stiffness. These values were tuned until overall gross behavior was observed to match that of the staged experimental dummy. The sequence can be seen in Figure 19 for IPTM crash test 4.[3] As expected, we found that one can optimize the Virtual CRASH settings until the overall simulated behavior is in good agreement with the observed behavior during staged tests.

**Conclusions**

We have tested Virtual CRASH for use in modeling pedestrian impacts. The simulator faithfully reproduces the expected behavior of projectiles during both the airborne and ground sliding phases of their trajectories. The throw distances as a function of impact speed behavior of the simulated dummy model does a good job reproducing the behavior observed during staged impact experiments as well as that which was observed in prior experiments.

**About the Authors**

Tony Becker has conducted extensive research in pedestrian and cyclist traffic crash investigation and published several articles and books that are widely used in the field. He is an ACTAR certified accident reconstructionist and trainer in Florida. He can be contacted at beckercrash@gmail.com.

Mike Reade, CD is the owner of Forensic Reconstruction Specialists Inc., a collision reconstruction consulting firm located in Moncton, New Brunswick Canada. He is also an Adjunct Instructor with the Institute of Police Technology and Management – University of North Florida (IPTM-UNF) based out of Jacksonville, Florida. He can be reached at reademw@gmail.com.

Bob Scurlock, Ph.D., ACTAR, is the owner Scurlock Scientific Services, LLC, an accident reconstruction consulting firm based out of Gainesville, Florida, USA. He is also a Research Associate at the University of Florida, Department of Physics. His website can be found at www.ScurlockPhD.com. He can be reached at BobScurlockPhD@gmail.com.

---

[3] A video of this simulation and staged collision footage can be seen online at:
https://youtu.be/FM9W9SCteYc

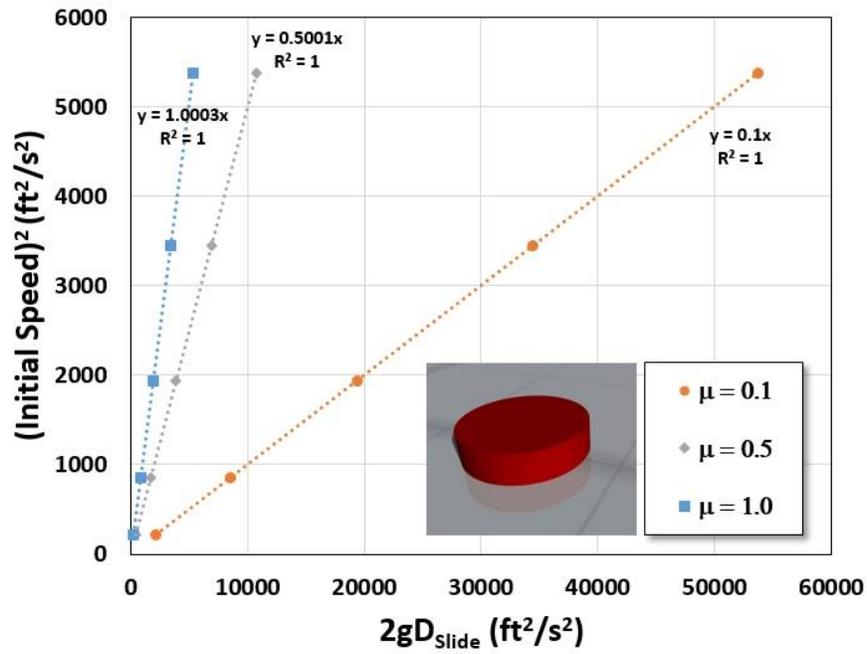

Figure 1: Results from puck slide-to-stop experiments in Virtual CRASH.

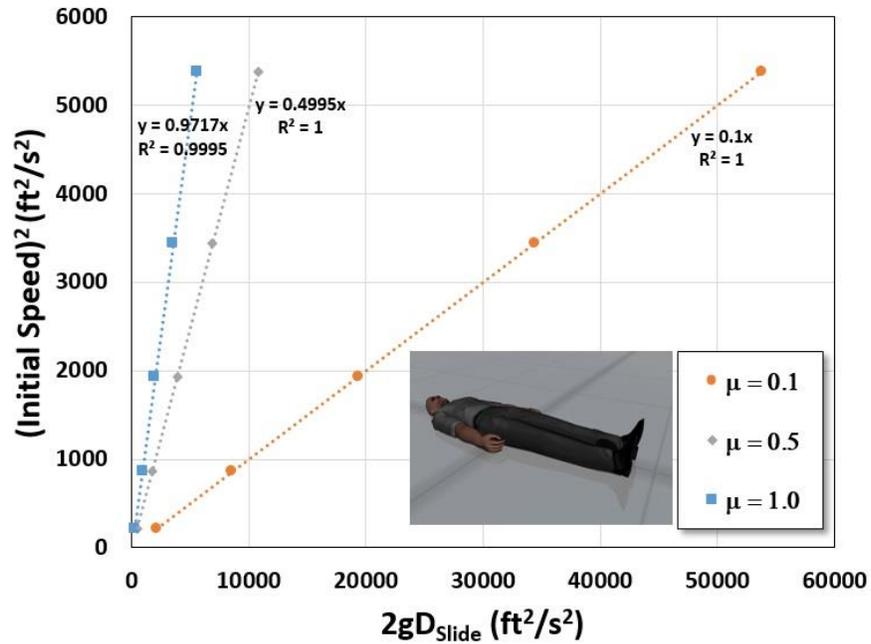

Figure 2: Results from dummy slide-to-stop experiments in Virtual CRASH.

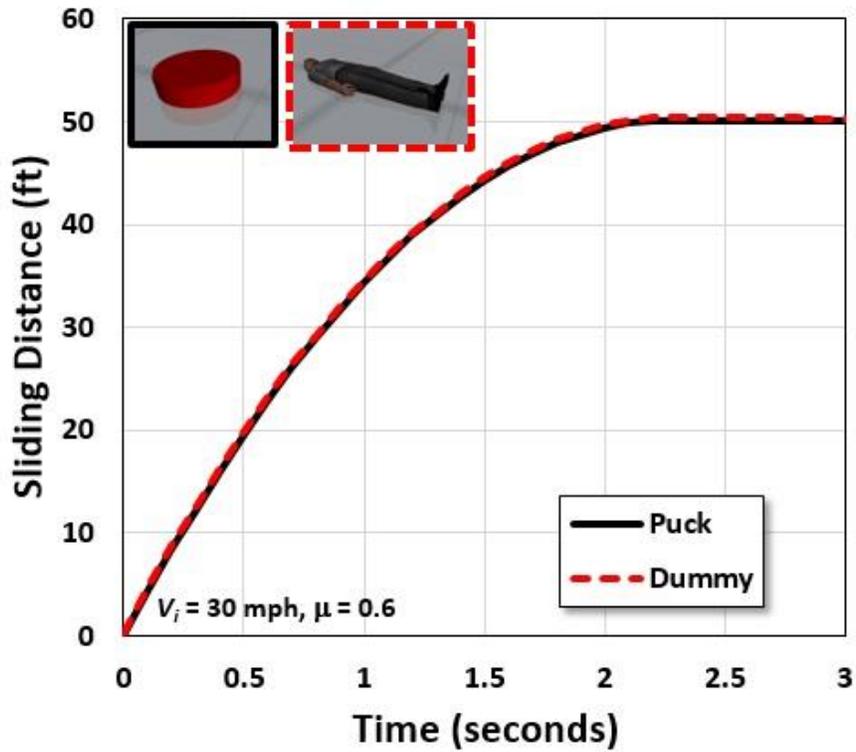

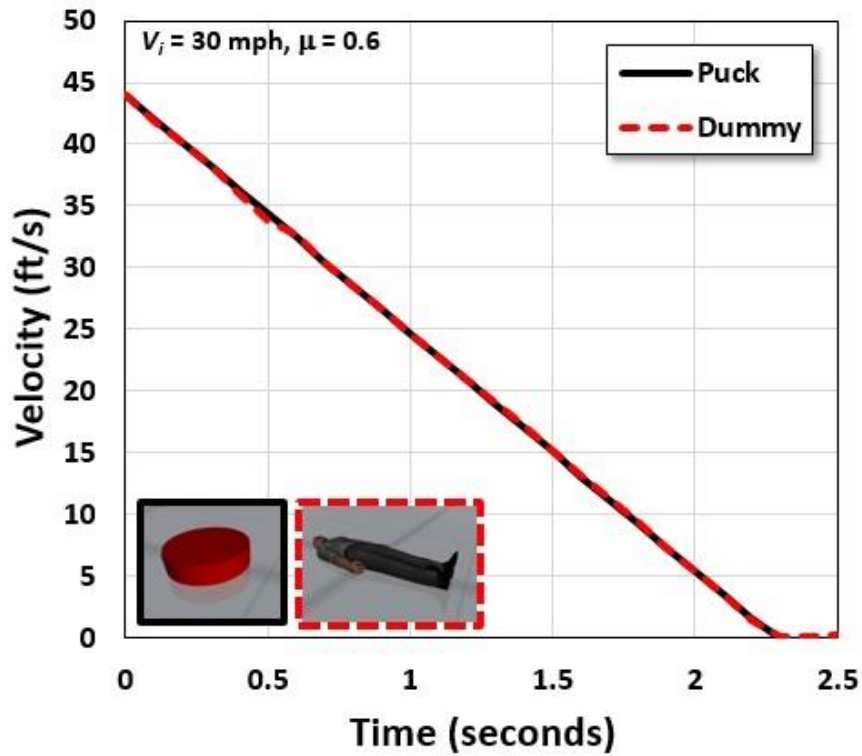

**Figure 3: Total slide distance versus time (top) and velocity versus time (bottom) for puck and dummy Virtual CRASH simulations. The results for the puck are shown in black. Results for the dummy are in red.**

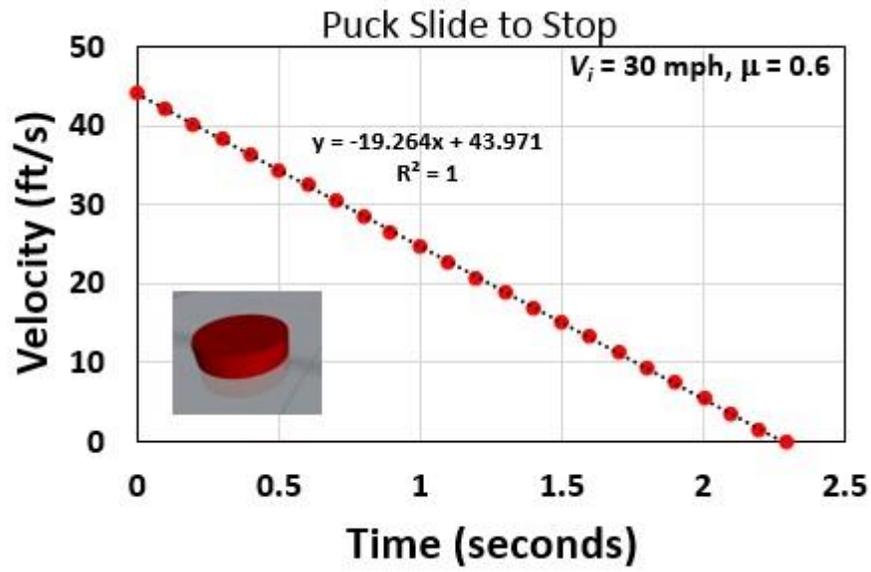

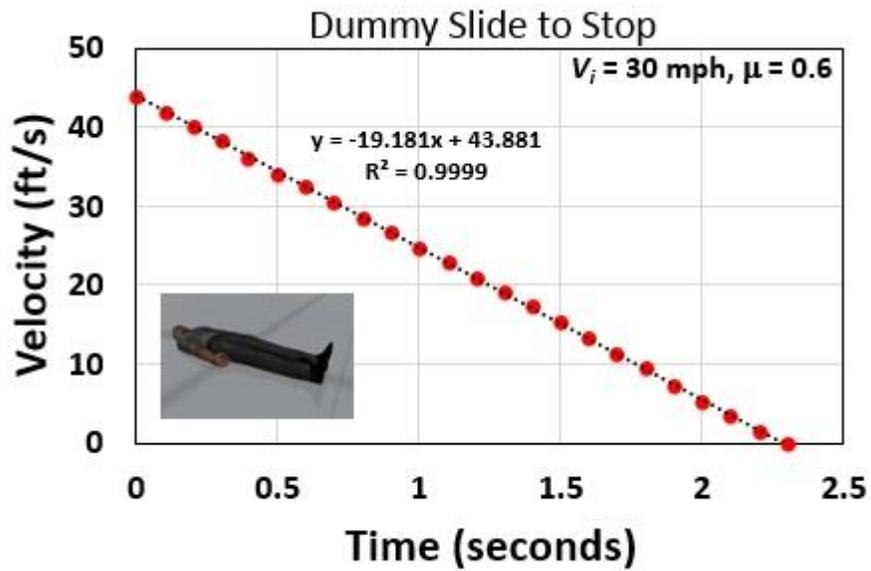

**Figure 4:** Velocity versus time for Puck (top) and Dummy (bottom) versus time. First-order polynomial fits yield the decelerations rates for the simulation.

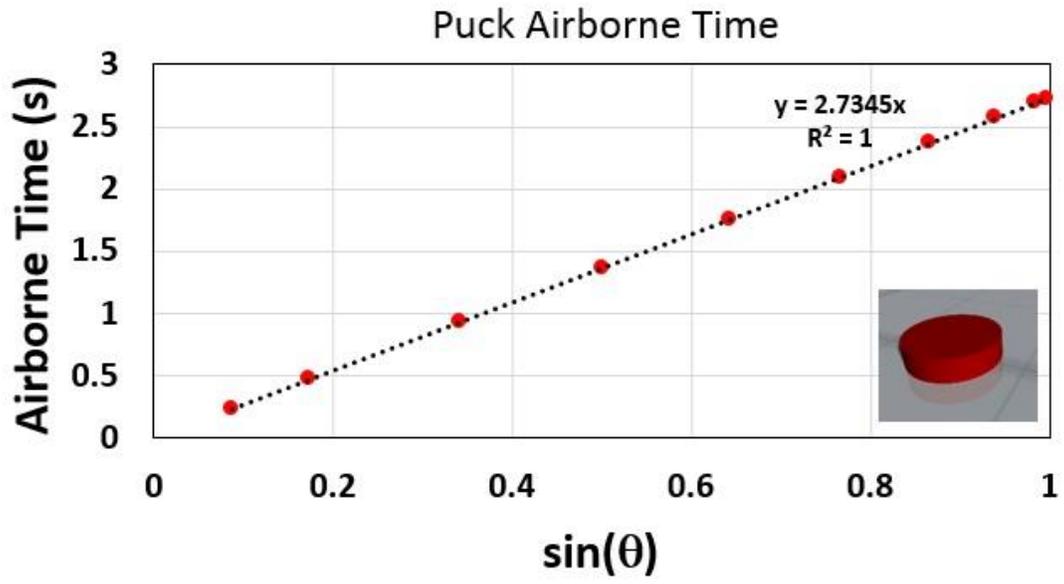

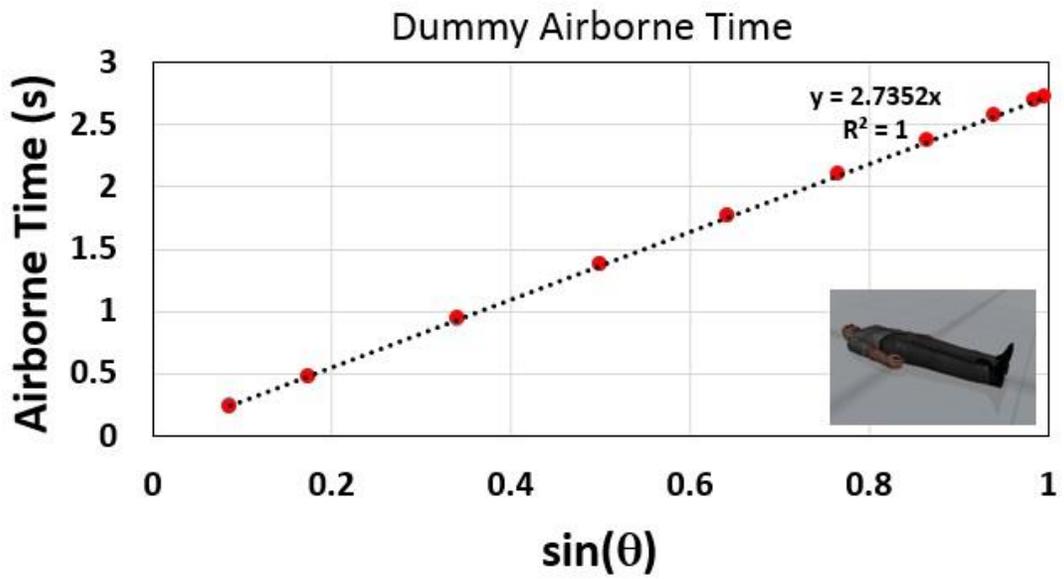

Figure 5: Total airborne time for puck (top) and dummy (bottom). First-order polynomial fits are shown.

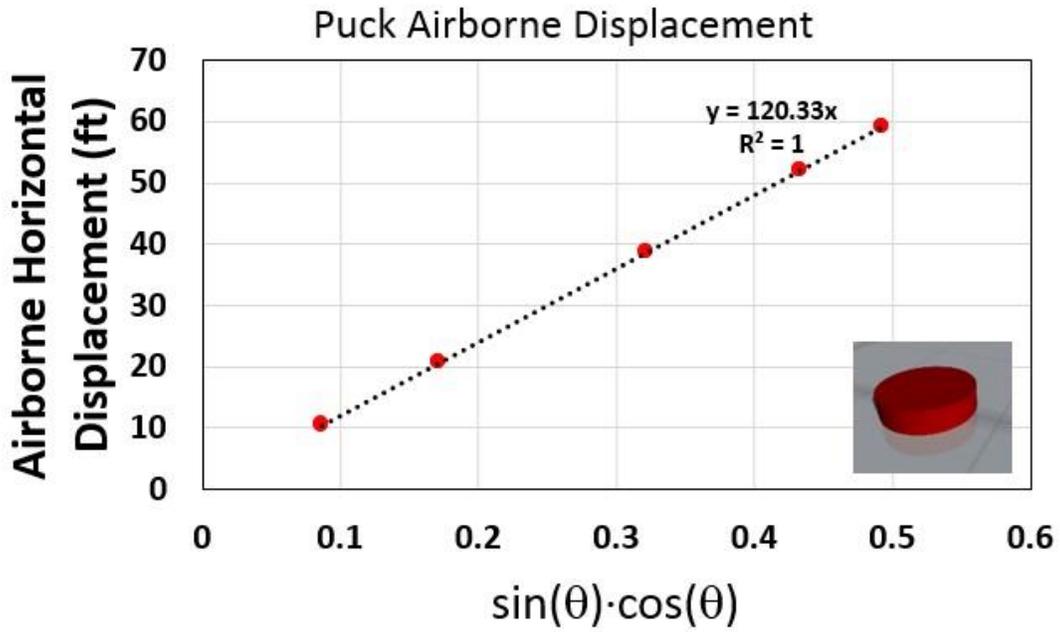

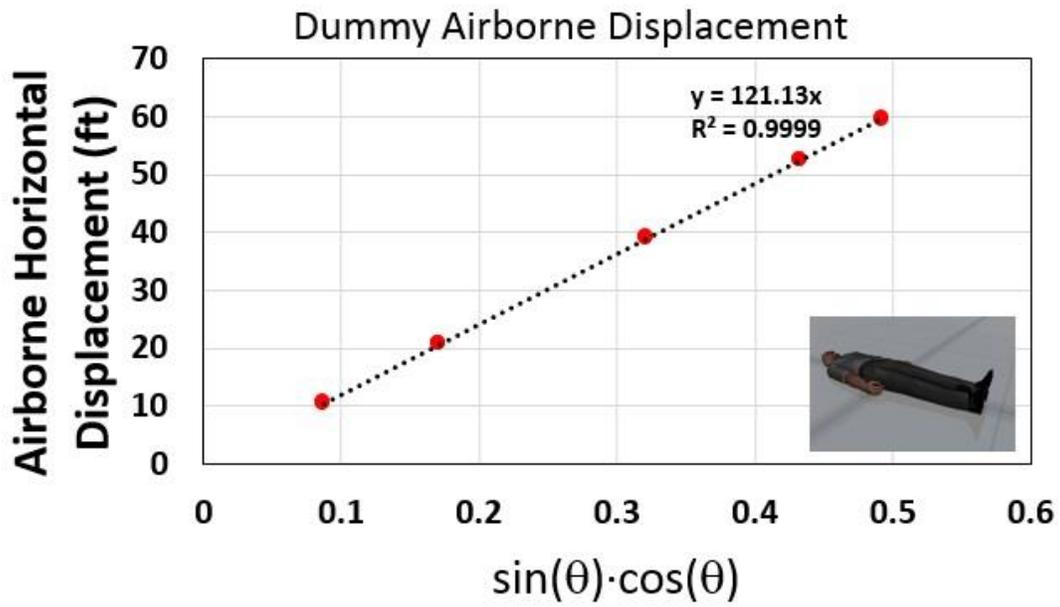

Figure 6: Total airborne horizontal displacement for puck (top) and dummy (bottom) as a function of $(\sin(\theta) \cdot \cos(\theta))$.

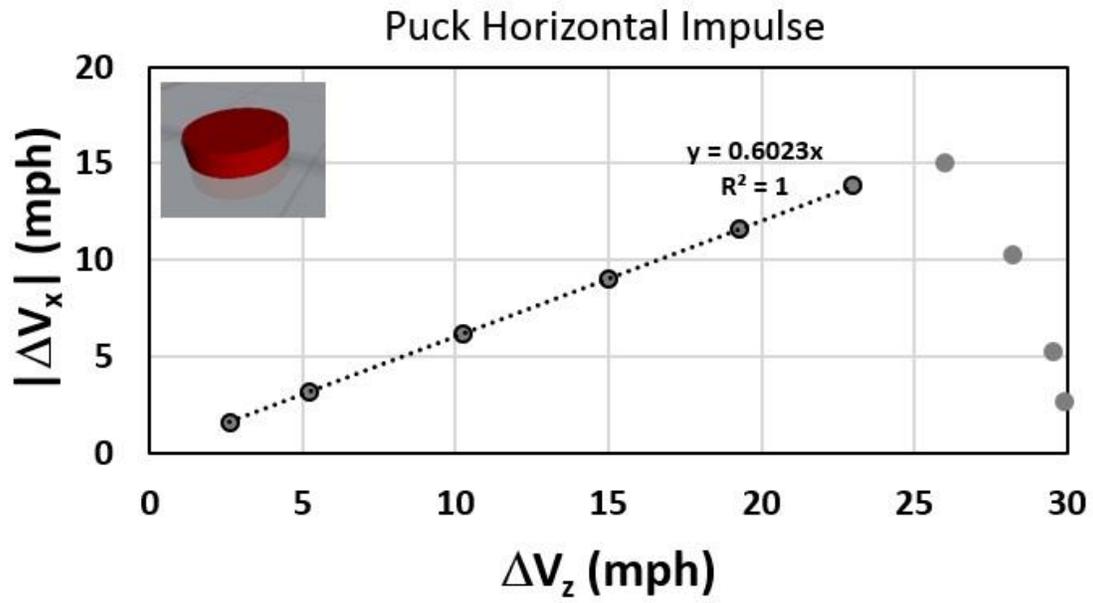

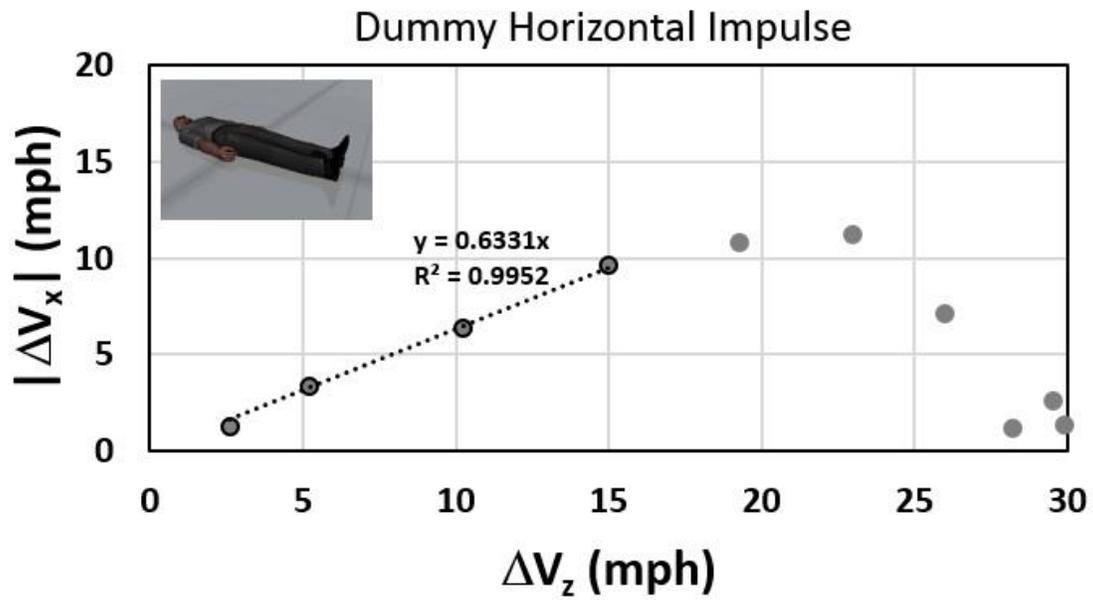

**Figure 7:** $\Delta v_x$ versus $\Delta v_z$ for the puck (top) and dummy (bottom) systems. First-order polynomial fits are performed to the linear region (open circles).

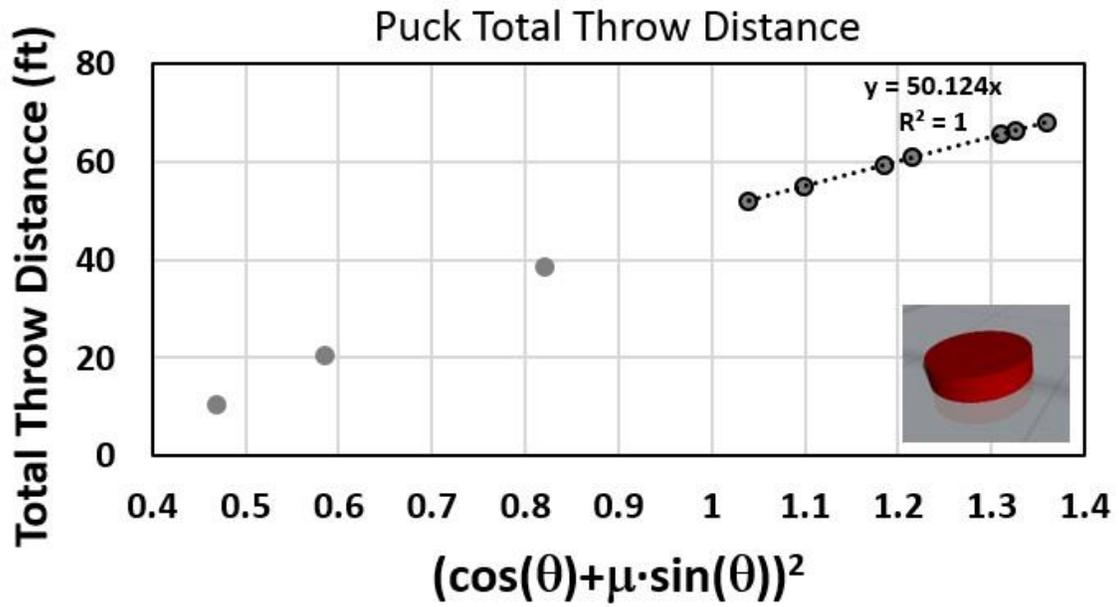

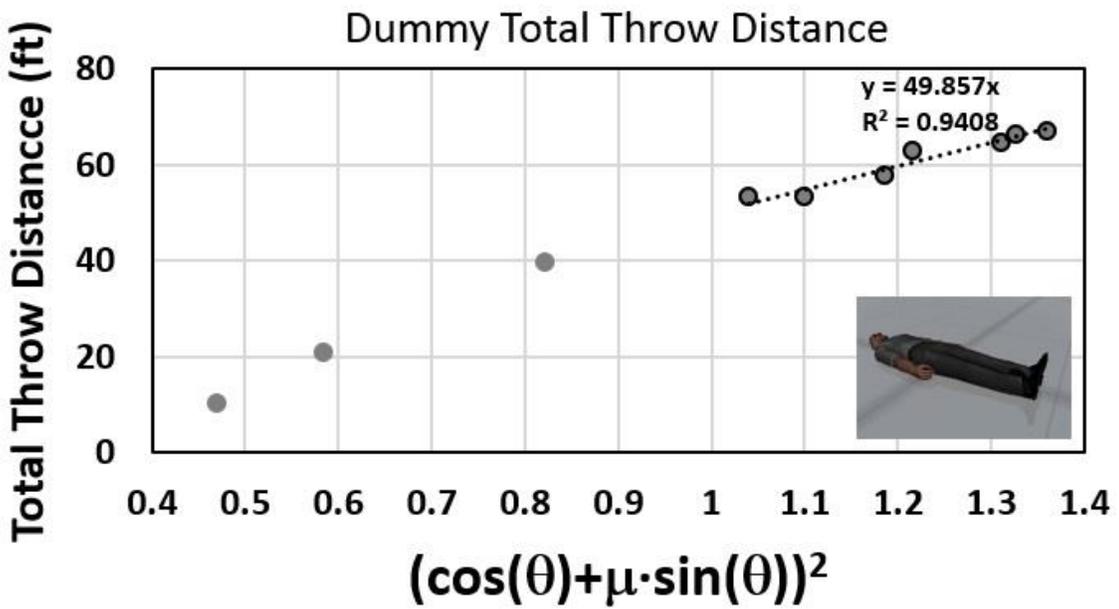

Figure 8: Total throw distance for puck (top) and dummy (bottom). First-order polynomial fits are performed to the linear region (open circles).

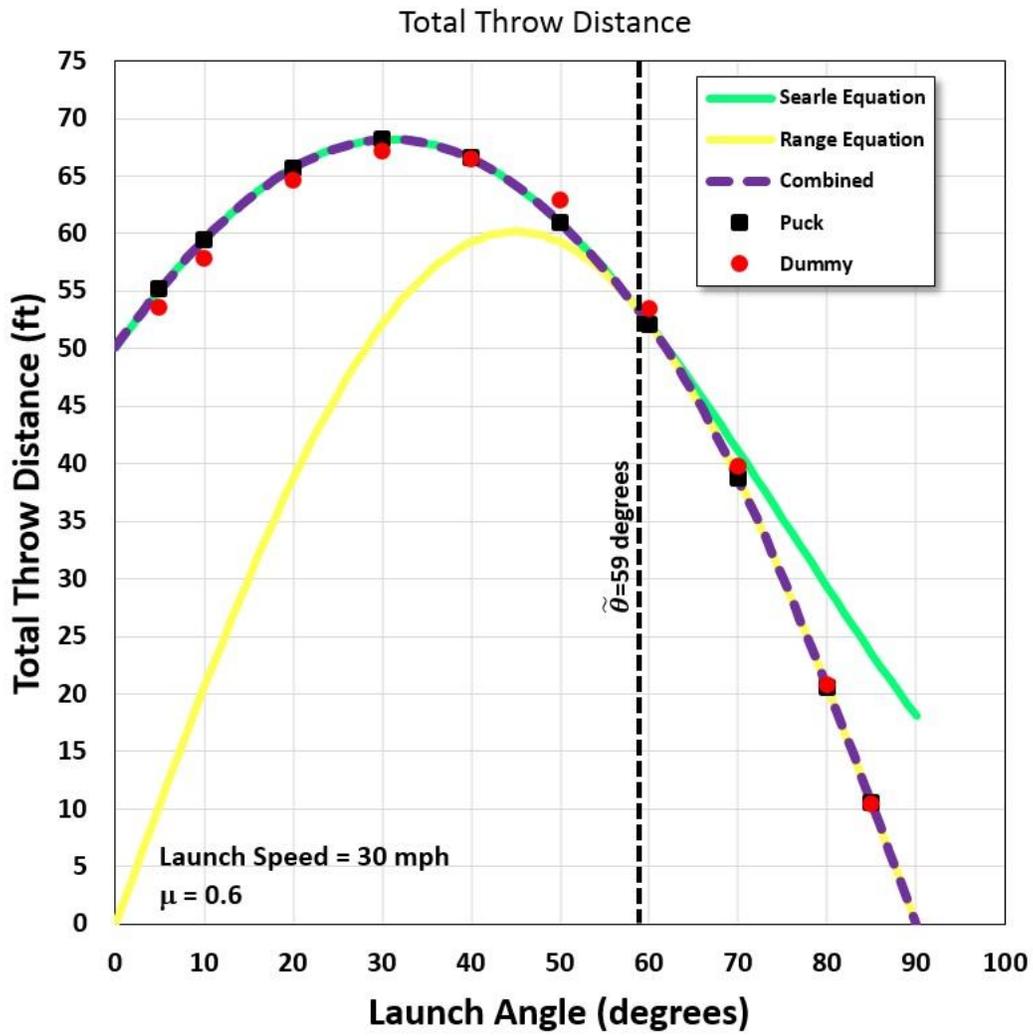

**Figure 9: Total throw distance as a function of launch angle for puck and dummy systems.**

| | PUCK | |
|---|---|---|
| θ (degrees) | Difference (ft) | %-Difference |
| 5 | 0.0193 | 0.035% |
| 10 | -0.0105 | -0.018% |
| 20 | -0.0981 | -0.149% |
| 30 | -0.0226 | -0.033% |
| 40 | -0.0136 | -0.020% |
| 50 | -0.0119 | -0.019% |
| 60 | -0.0042 | -0.008% |
| 70 | -0.0023 | -0.006% |
| 80 | -0.0003 | -0.002% |
| 85 | 0.0391 | 0.374% |
| Average | -0.0105 | 0.015% |
| Standard Deviation | 0.0356 | 0.135% |
| Minimum | -0.0981 | -0.149% |
| Maximum | 0.0391 | 0.374% |

| | DUMMY | |
|---|---|---|
| θ (degrees) | Difference (ft) | %-Difference |
| 5 | -1.61 | -2.92% |
| 10 | -1.71 | -2.87% |
| 20 | -1.08 | -1.65% |
| 30 | -1.06 | -1.55% |
| 40 | -0.03 | -0.04% |
| 50 | 1.90 | 3.12% |
| 60 | 1.34 | 2.56% |
| 70 | 1.08 | 2.79% |
| 80 | 0.22 | 1.05% |
| 85 | -0.02 | -0.21% |
| Average | -0.10 | 0.03% |
| Standard Deviation | 1.26 | 2.29% |
| Minimum | -1.71 | -2.92% |
| Maximum | 1.90 | 3.12% |

**Table 1: Difference between analytic solution and simulation for the puck (top) and dummy (bottom) systems.**

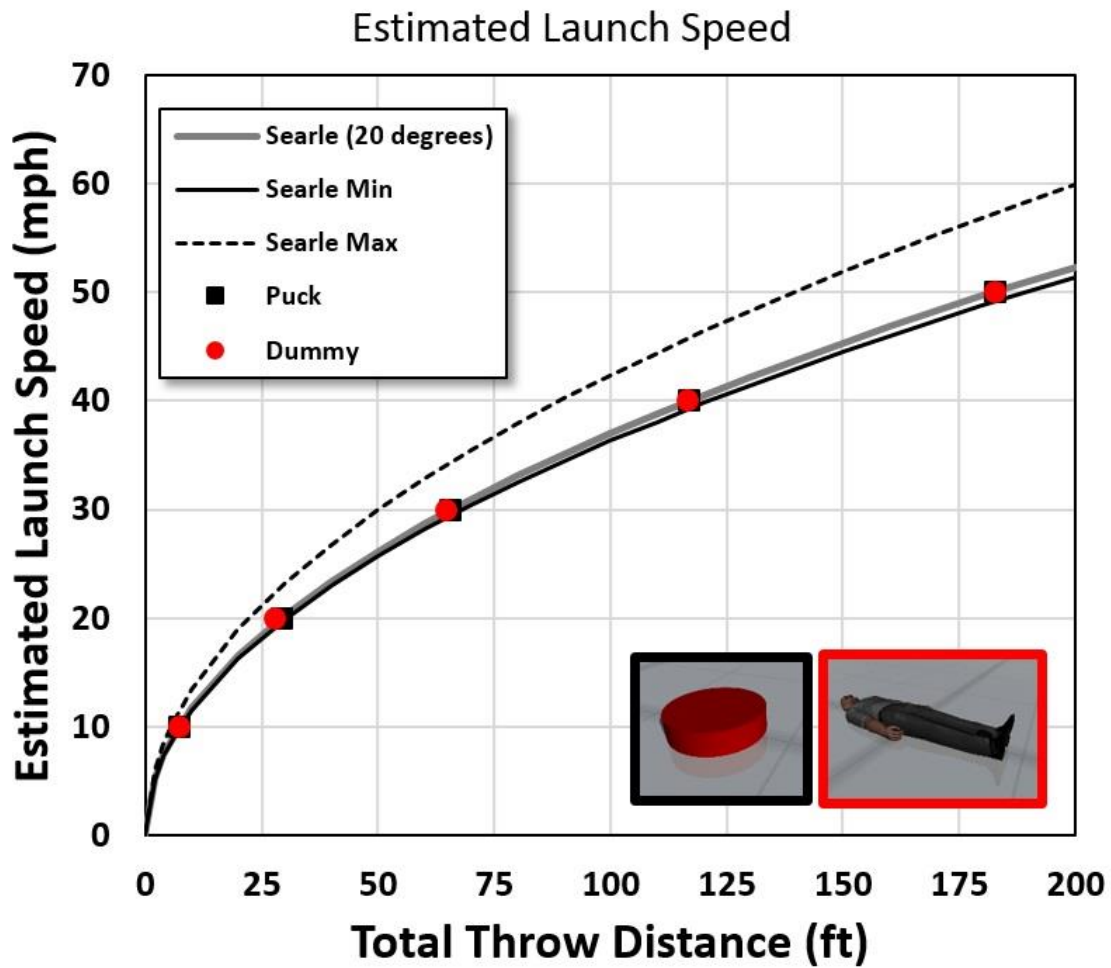

**Figure 10: Launch speed estimates as a function total throw distance. Results from puck and dummy simulations are shown for a 20 degree launch.**

| PUCK | | |
|---|---|---|
| Launch Speed (mph) | Difference (mph) | %-Difference |
| 10 | 0.51 | 5.07% |
| 20 | 0.01 | 0.04% |
| 30 | -0.57 | -1.88% |
| 40 | -0.24 | -0.60% |
| 50 | 1.92 | 3.85% |
| Average | 0.33 | 1.30% |
| Standard Deviation | 0.98 | 3.00% |
| Minimum | -0.57 | -1.88% |
| Maximum | 1.92 | 5.07% |

| DUMMY | | |
|---|---|---|
| Launch Speed (mph) | Difference (mph) | %-Difference |
| 10 | 0.42 | 4.21% |
| 20 | -0.46 | -2.30% |
| 30 | -0.76 | -2.53% |
| 40 | -0.30 | -0.75% |
| 50 | 1.93 | 3.85% |
| Average | 0.17 | 0.50% |
| Standard Deviation | 1.08 | 3.30% |
| Minimum | -0.76 | -2.53% |
| Maximum | 1.93 | 4.21% |

**Table 2: Differences between analytic solution of estimated launch speed and simulations of puck and dummy systems for a 20 degree launch.**

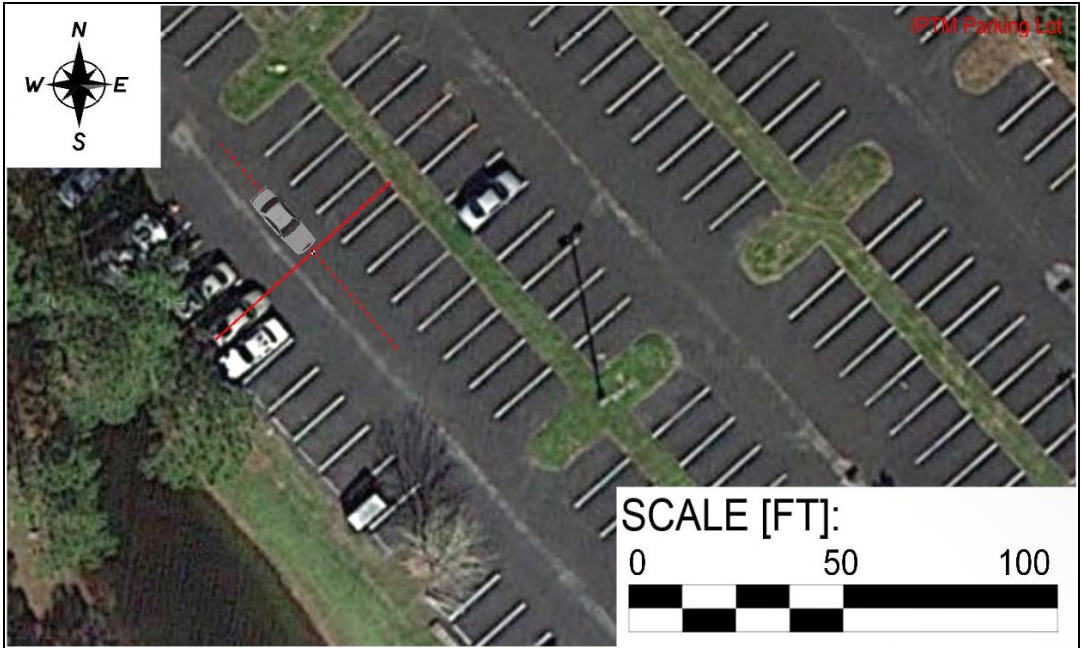
**Figure 11: Aerial view of test site at IPTM.**

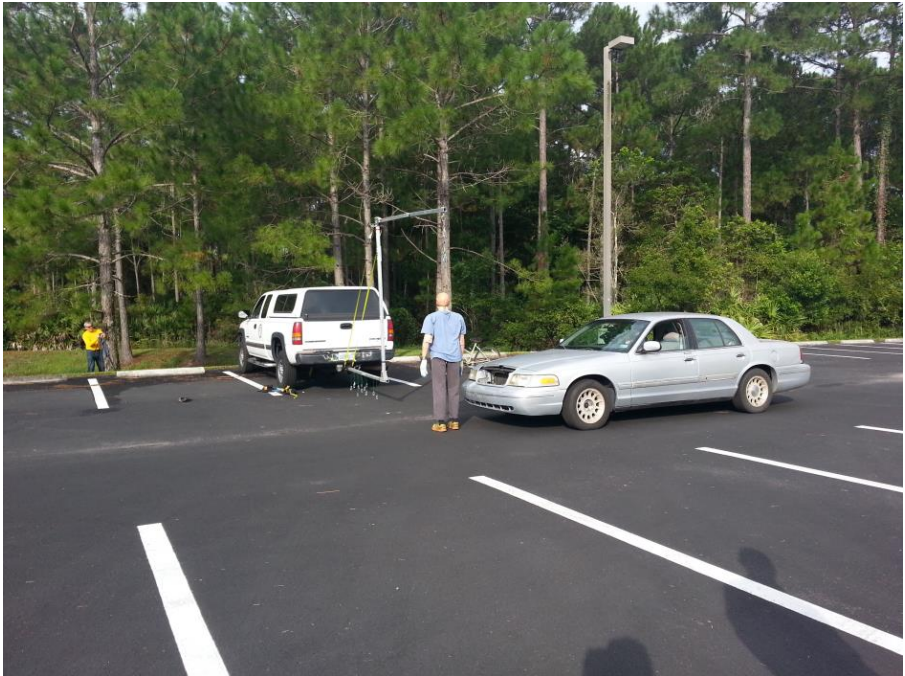
**Figure 12: Anthrophonic test dummy pre-impact configuration.**

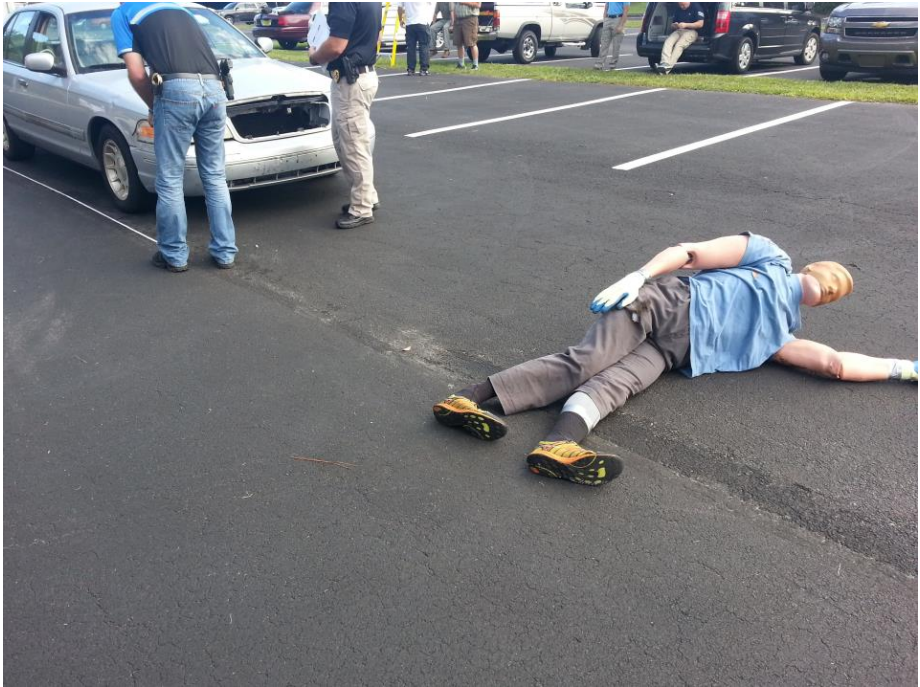
**Figure 13: Dummy post-impact position.**

| Test Number | Impact Location on Dummy | Total Throw Distance (ft) | Impact Speed (mph) |
|---|---|---|---|
| \multicolumn{4}{|c|}{August 12, 2015 IPTM Staged Impact Data} |
| 1 | Right | 34.8 | 23.9 |
| 2 | Right | 37.2 | 24.4 |
| 3 | Right | 48.5 | 27.1 |
| 4 | Right | 29.7 | 21.2 |
| 5 | Rear | 57.5 | 34.2 |
| 6 | Front | 81.2 | 38.0 |
| 7 | Left | 85.4 | 40.3 |

Table 3: Summary of staged collision data.

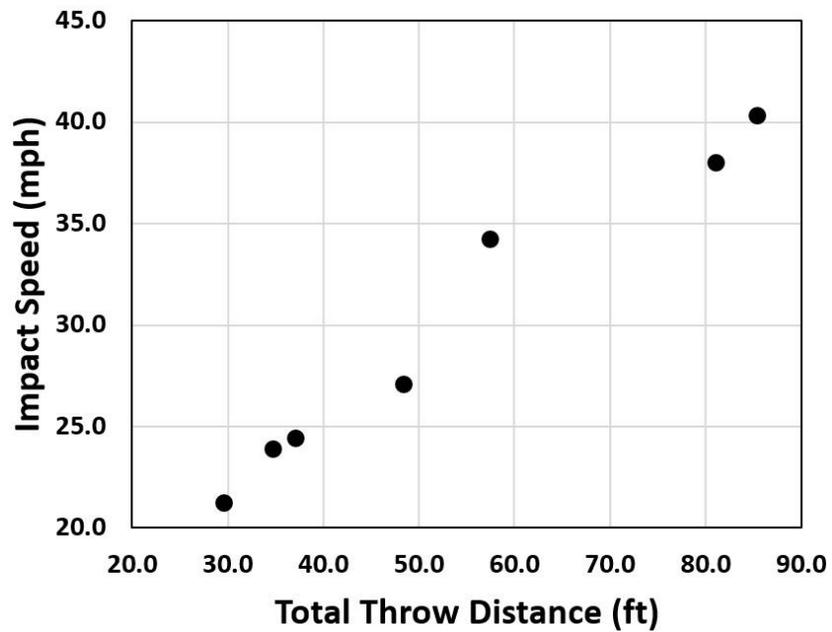

Figure 14: Plot depicting impact speed versus total throw distance for staged impacts.

| Virtual Crash Simulation Settings | | | | |
|---|---|---|---|---|
| Simulation Parameter | Scenario 1 | Scenario 2 | Scenario 3 | Scenario 4 |
| Crown Vic Lateral Offset (ft) | 0.5 | 0.5 | 1.5 | 1.5 |
| Dummy Orientation (degrees) | 0 | 90 | 0 | 90 |
| Ground Contact coefficient-of-friction | 0.511 | 0.511 | 0.511 | 0.511 |
| Vehicle Contact coefficient-of-friction | 0.5 | 0.5 | 0.5 | 0.5 |
| Ground Contact coefficient-of-restitution | 0 | 0 | 0 | 0 |
| Vehicle Contact coefficient-of-restitution | 0 | 0 | 0 | 0 |
| Crown Vic Total Total Weight (+driver) (lbs) | 4128 | 4128 | 4128 | 4128 |
| Crown Vic Average Decelaration Rate (g) | 0.73 | 0.73 | 0.73 | 0.73 |
| Dummy Weight (lbs) | 49 | 49 | 49 | 49 |
| Dummy Height (ft) | 5.83 | 5.83 | 5.83 | 5.83 |
| Dummy Joint Stiffness (N/ft) | 0.61 | 0.61 | 0.61 | 0.61 |

Table 4: Summary of simulation input parameter settings used in Virtual CRASH 3.0.

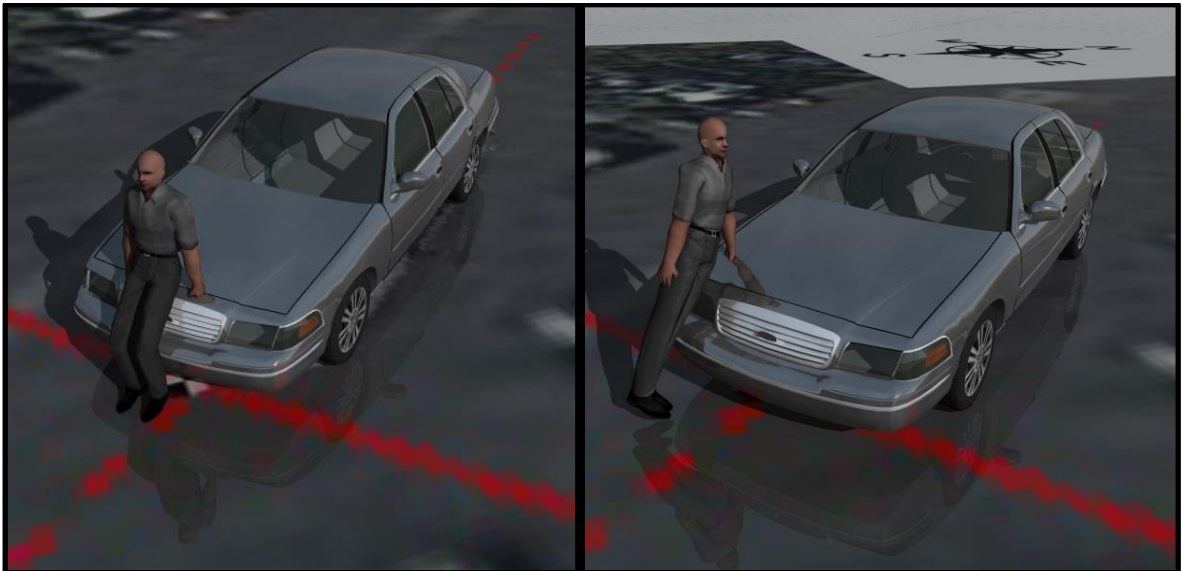

Figure 15: Moment-of-impact depicted for Scenario 1 (left) and Scenario 4 (right).

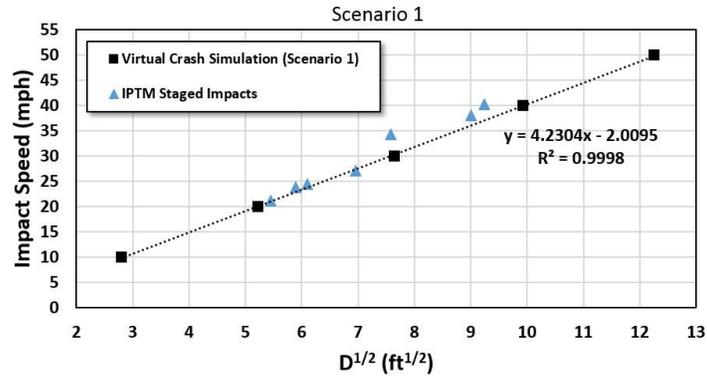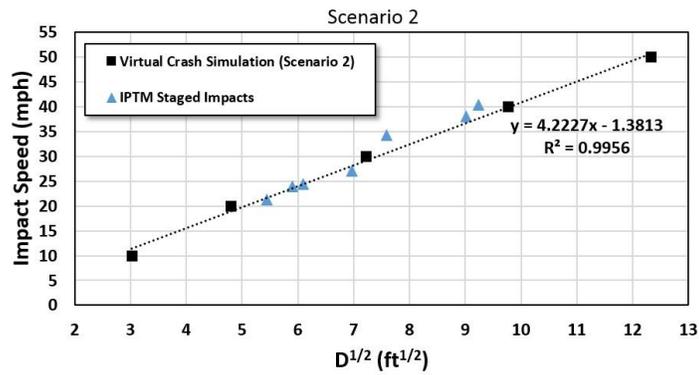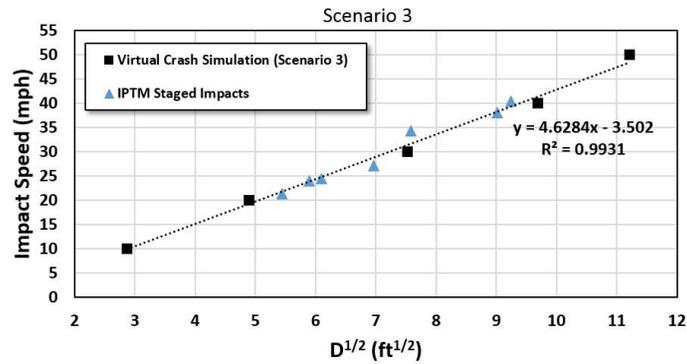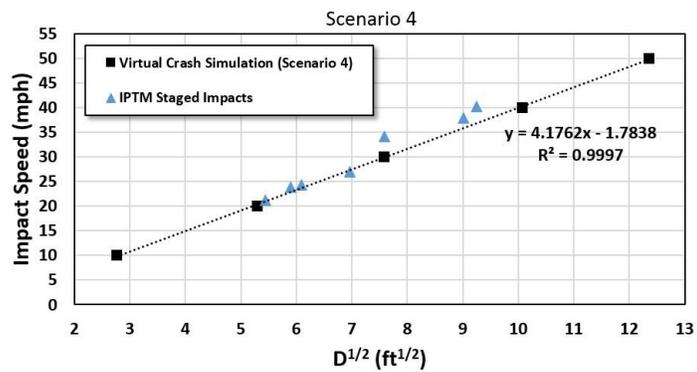

Figure 16: Plots depicting relationship between Impact Speed and the square root of the total throw distance. Plots are shown for the four simulated scenarios. The IPTM test data is plotted as well.

| Test Number | IPTM Total Throw Distance (ft) | IPTM Impact Speed (mph) | Scenario 1 Fit (mph) | Scenario 1 Fit Residual (Obs - Fit) (mph) | Scenario 1 Fit (Residual/Obs) (%) |
|---|---|---|---|---|---|
| 1 | 34.8 | 23.9 | 22.7 | 1.18 | 4.94% |
| 2 | 37.2 | 24.4 | 23.5 | 0.86 | 3.51% |
| 3 | 48.5 | 27.1 | 27.2 | -0.12 | -0.43% |
| 4 | 29.7 | 21.2 | 20.8 | 0.38 | 1.79% |
| 5 | 57.5 | 34.2 | 29.8 | 4.47 | 13.06% |
| 6 | 81.2 | 38.0 | 35.7 | 2.29 | 6.02% |
| 7 | 85.4 | 40.3 | 36.7 | 3.62 | 8.97% |
| | | | Average | 1.81 | 5.41% |
| | | | Standard Deviation | 1.71 | 4.52% |
| | | | Minimum | -0.12 | -0.43% |
| | | | Maximum | 4.47 | 13.06% |

| Test Number | IPTM Total Throw Distance (ft) | IPTM Impact Speed (mph) | Scenario 2 Fit (mph) | Scenario 2 Fit Residual (Obs - Fit) (mph) | Scenario 2 Fit (Residual/Obs) (%) |
|---|---|---|---|---|---|
| 1 | 34.8 | 23.9 | 23.4 | 0.46 | 1.94% |
| 2 | 37.2 | 24.4 | 24.3 | 0.11 | 0.45% |
| 3 | 48.5 | 27.1 | 28.1 | -1.00 | -3.71% |
| 4 | 29.7 | 21.2 | 21.5 | -0.26 | -1.24% |
| 5 | 57.5 | 34.2 | 30.7 | 3.48 | 10.18% |
| 6 | 81.2 | 38.0 | 36.9 | 1.08 | 2.84% |
| 7 | 85.4 | 40.3 | 37.9 | 2.37 | 5.88% |
| | | | Average | 0.89 | 2.33% |
| | | | Standard Deviation | 1.56 | 4.61% |
| | | | Minimum | -1.00 | -3.71% |
| | | | Maximum | 3.48 | 10.18% |

Table 5: Summary of results for simulation Scenarios 1 (top) and 2 (bottom).

| Test Number | IPTM Total Throw Distance (ft) | IPTM Impact Speed (mph) | Scenario 3 Fit (mph) | Scenario 3 Fit Residual (Obs - Fit) (mph) | Scenario 3 Fit (Residual/Obs) (%) |
|---|---|---|---|---|---|
| 1 | 34.8 | 23.9 | 24.2 | -0.35 | -1.48% |
| 2 | 37.2 | 24.4 | 25.1 | -0.70 | -2.87% |
| 3 | 48.5 | 27.1 | 28.8 | -1.79 | -6.61% |
| 4 | 29.7 | 21.2 | 22.3 | -1.09 | -5.16% |
| 5 | 57.5 | 34.2 | 31.5 | 2.72 | 7.95% |
| 6 | 81.2 | 38.0 | 37.6 | 0.36 | 0.94% |
| 7 | 85.4 | 40.3 | 38.6 | 1.65 | 4.10% |
| | | Average | | 0.11 | -0.45% |
| | | Standard Deviation | | 1.59 | 5.17% |
| | | Minimum | | -1.79 | -6.61% |
| | | Maximum | | 2.72 | 7.95% |

| Test Number | IPTM Total Throw Distance (ft) | IPTM Impact Speed (mph) | Scenario 4 Fit (mph) | Scenario 4 Fit Residual (Obs - Fit) (mph) | Scenario 4 Fit (Residual/Obs) (%) |
|---|---|---|---|---|---|
| 1 | 34.8 | 23.9 | 23.5 | 0.41 | 1.73% |
| 2 | 37.2 | 24.4 | 24.3 | 0.08 | 0.34% |
| 3 | 48.5 | 27.1 | 28.0 | -0.93 | -3.44% |
| 4 | 29.7 | 21.2 | 21.6 | -0.36 | -1.72% |
| 5 | 57.5 | 34.2 | 30.6 | 3.63 | 10.60% |
| 6 | 81.2 | 38.0 | 36.6 | 1.38 | 3.64% |
| 7 | 85.4 | 40.3 | 37.6 | 2.70 | 6.70% |
| | | Average | | 0.99 | 2.55% |
| | | Standard Deviation | | 1.67 | 4.88% |
| | | Minimum | | -0.93 | -3.44% |
| | | Maximum | | 3.63 | 10.60% |

Table 6: Summary of results for simulation Scenarios 3 (top) and 4 (bottom).

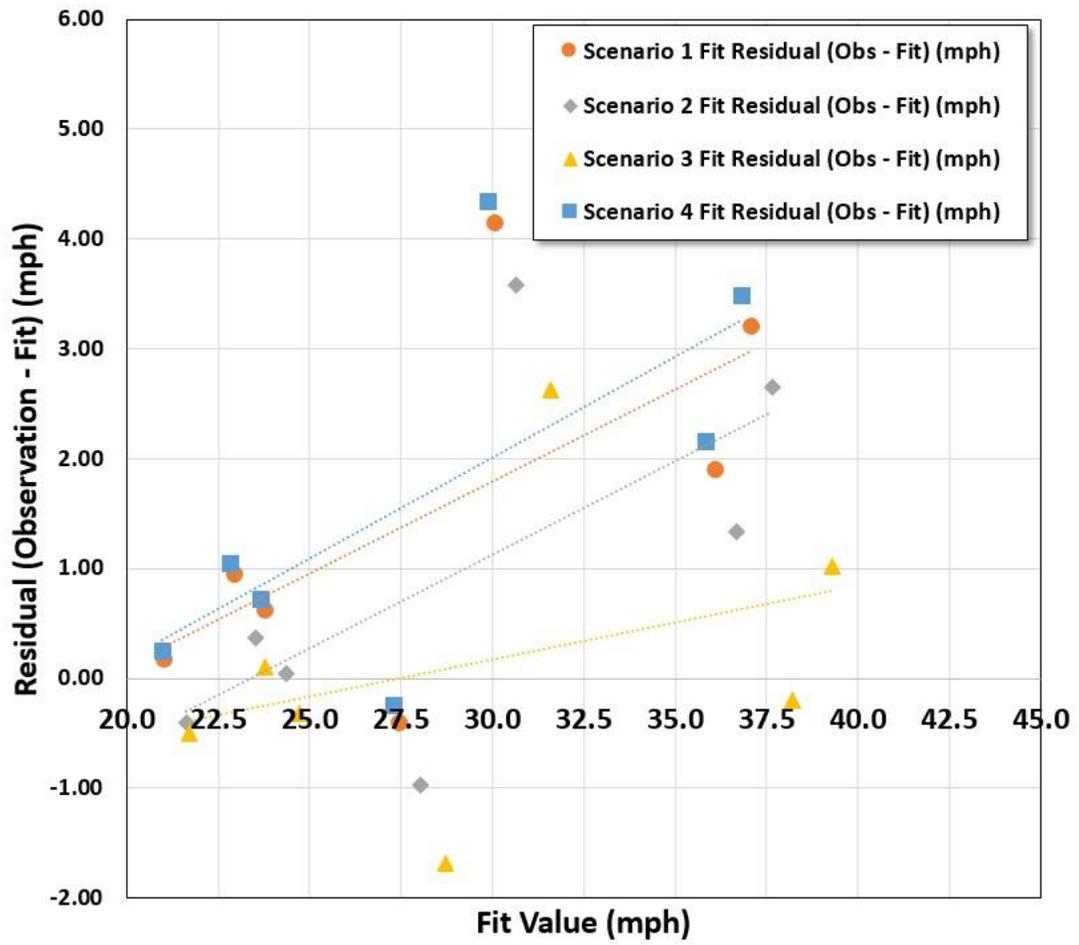

Figure 17: Plot of residuals as a function of simulation fit values for all four Scenarios.

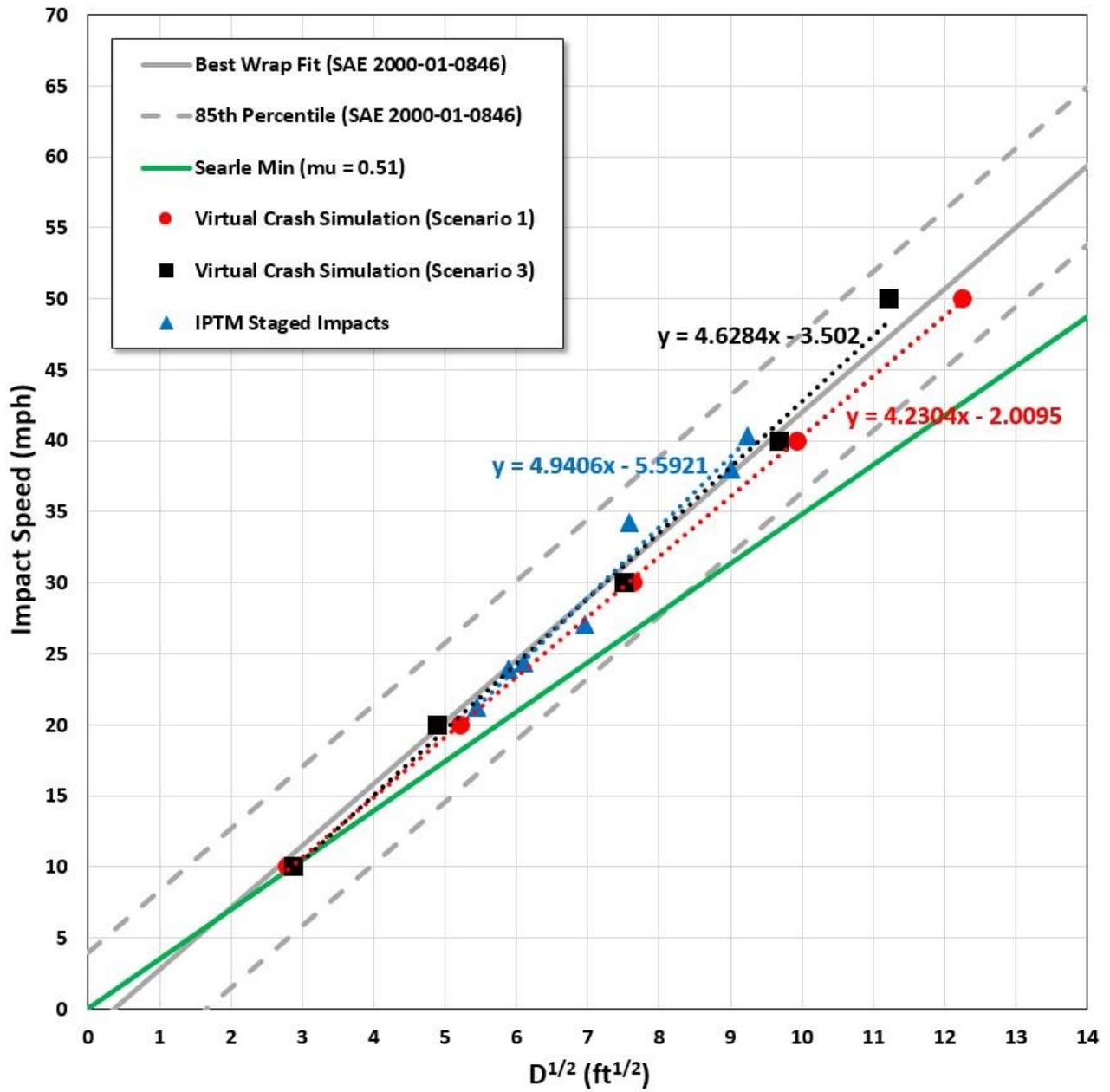

Figure 18: Plot depicting impact speed versus square root of total throw distance for Scenarios 1 and 3.

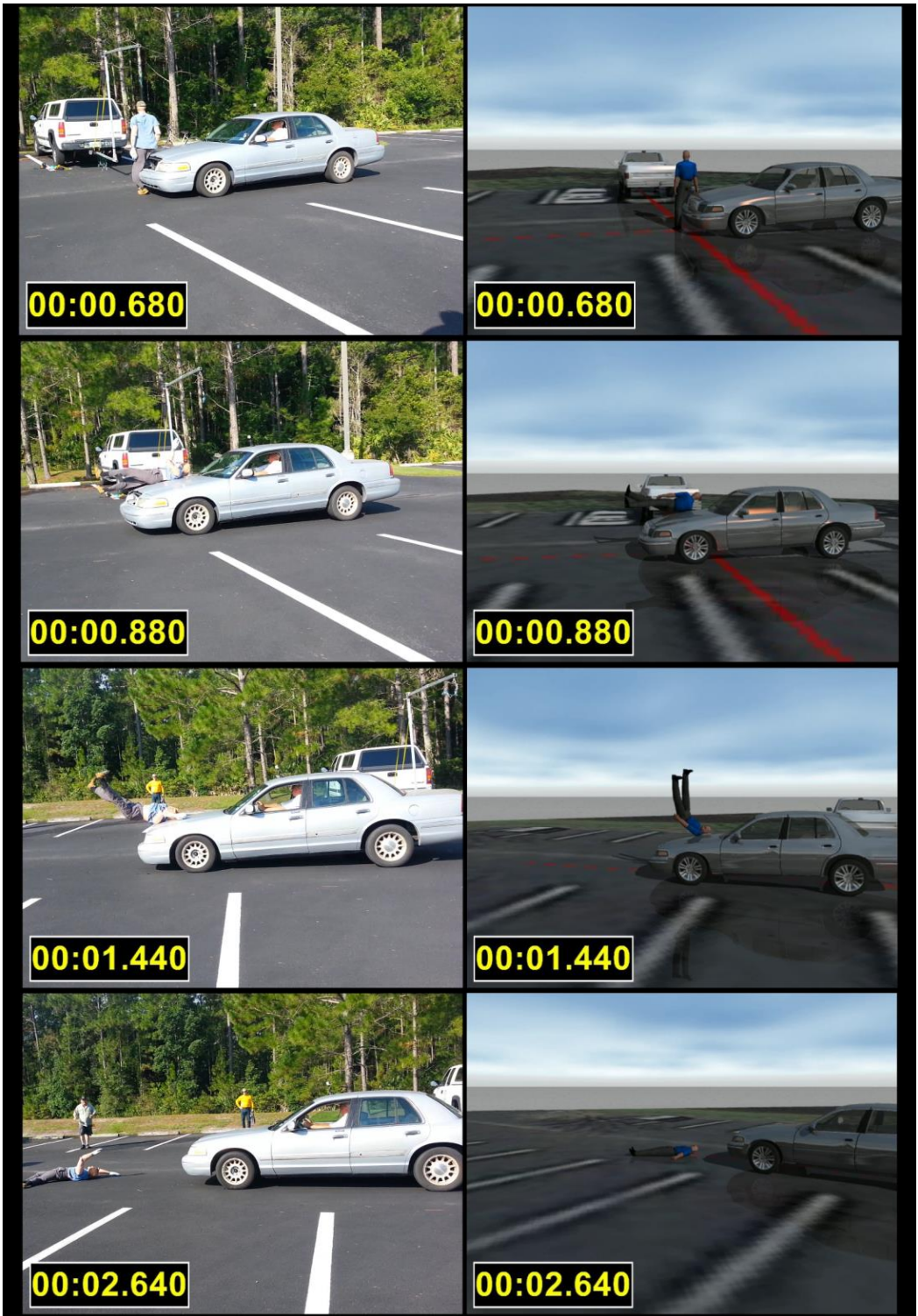

**Figure 19: Comparing simulation of Test #4 of IPTM staged impact.**